\RequirePackage{fix-cm}
\documentclass[smallextended]{svjour3}       
\smartqed  
\usepackage{graphicx}
\usepackage{url}
\usepackage[resetlabels,labeled]{multibib}
\newcites{I}
{References to ICSE distinguished papers}  
 \usepackage{multirow}

\usepackage{enumerate}

\usepackage{color}
\usepackage{ifthen}
\newboolean{showcomments}
\setboolean{showcomments}{false} 
\ifthenelse{\boolean{showcomments}}
    {
        \newcommand{\emelie}[1]{\textcolor{red}{{\it [Emelie says: #1]}}}
        \newcommand{\teresa}[1]{\textcolor{red}{{\it [Teresa says: #1]}}}
        \newcommand{\peggy}[1]{\textcolor{blue}{{\it [Peggy says: #1]}}}
        \newcommand{\per}[1]{\textcolor{cyan}{{\it [Per says: #1]}}}
        \newcommand{\martin}[1]{\textcolor{cyan}{{\it [Martin says: #1]}}}
        \newcommand{\cassie}[1]{\textcolor{cyan}{{\it [Cassie says: #1]}}}
        \newcommand{\othercomment}[1]{\textcolor{magenta}{{\it [#1]}}}
    }
    {
        \newcommand{\emelie}[1]{}
        \newcommand{\teresa}[1]{}

        \newcommand{\peggy}[1]{}
        \newcommand{\per}[1]{}
        \newcommand{\cassie}[1]{}
        \newcommand{\martin}[1]{}
        \newcommand{\othercomment}[1]{}
    }

\begin{document}

\title{
How software engineering research aligns with design science: A review
}


\author{Emelie Engstr\"om         \and
        Margaret-Anne Storey \and
        Per Runeson \and
        Martin H\"ost \and
        Maria Teresa Baldassarre
}


\institute{Emelie Engstr\"om, Per Runeson,  Martin H\"ost  \at
              Lund University, Sweden\\
              \email{[emelie.engstrom,per.runeson,martin.host]@cs.lth.se}           
           \and
          Margaret-Anne Storey  \at
             University of Victoria, Canada\\
             \email{mstorey@uvic.ca}
           \and 
           Maria Teresa Baldassarre  \at
               University of Bari, Italy\\
              \email{mariateresa.baldassarre@uniba.it} 
}

\date{Received: date / Accepted: date}

\maketitle

\begin{abstract}
\emph{Background:} Assessing and communicating software engineering research can be challenging.
 Design science is recognized as an appropriate research paradigm for applied research, but is seldom referred to in software engineering.
 Applying the design science lens to software engineering research may improve the assessment and communication of research contributions. 
\emph{Aim:} The aim of this study is 1) to understand whether the design science lens helps summarize and assess software engineering research contributions, and 2) to characterize different types of design science contributions in the software engineering literature.   
\emph{Method:} In previous research, we developed a visual abstract template, summarizing the core constructs of the design science paradigm. 
 In this study, we use this template in a review of a set of 38 top software engineering publications to extract and analyze their design science contributions. 
\emph{Results:} We identified five clusters of papers, classifying them according to their alignment with the design science paradigm. 
\emph{Conclusions:} The design science lens helps emphasize the theoretical contribution of research output---in terms of technological rules---and reflect on the practical relevance, novelty and rigor of the rules proposed by the research.


\end{abstract}

\section{Introduction}
\label{sec:introduction}

Design science is a paradigm for conducting and communicating applied research such as software engineering. Similar to other design sciences, much software engineering research aims to design solutions to practical problems in a real-world context. The goal of design science research is to produce prescriptive knowledge for professionals in a discipline and to share empirical insights gained from investigations of the prescriptions applied in context~\cite{aken_management_2004}. Such knowledge is referred to as ``design knowledge'' as it helps practitioners design solutions to their problems. 

Design science is an established research paradigm\footnote{By paradigm, we refer to van Aken's definition: ``the combination of research questions asked, the research methodologies allowed to answer them and the nature of the pursued research products''~\cite{van_aken_management_2005}.}  in the fields of information systems~\cite{hevner_design_2004} and other engineering disciplines, such as mechanical, civil, architectural, and manufacturing engineering\footnote{springer.com/journal/163 Research in Engineering Design}. %
It is also increasingly used in computer science; for example, it is now accepted as the \emph{de facto} paradigm for presenting design contributions from information visualization research~\cite{sedlmair2012}.
Although Wierenga has promoted design science for capturing design knowledge in software engineering~\cite{wieringa_what_2014}, we seldom see it being referred to in our field (although there are some exceptions~\cite{Wohlin2015}). 
We are puzzled by its low adoption as the use of this lens could increase the clarity of research contributions for both practitioners and researchers, as it has been shown to do in other fields~\cite{Shneiderman2016}. 

The goal of our research is to investigate if and how the design science paradigm may be a viable way to assess and communicate research contributions in existing software engineering literature. 
To this end, 
we consider a set of software engineering research papers and view these contributions through a design science lens 
by using and improving a visual abstract template we previously developed to showcase design knowledge~\cite{StoreyESEM17}.

We inspected 38 ACM distinguished papers published at the International Conference on Software Engineering (ICSE) over a five-year period---publications considered by many in the community as well-known exemplars of fine software engineering research, and papers that are expected to broadly represent the diverse topics addressed by our research community.
Although these papers set a high bar for framing their research contributions, 
we found that the design science lens improved our understanding of their contributions. 
Also,  
most of the papers described research contributions that are congruent with the design science paradigm, even though none of them explicitly used the term. 
Applying this lens 
helped us elucidate certain aspects of the contributions (such as relevance, novelty and rigor), which in some cases were obscured by the original framing of the paper.
However, not all the papers we considered produced design knowledge, thus some research publications do not benefit from using this lens.   

Our analysis from this exercise led to five clusters of papers 
based on the type of design knowledge reported. We compare the papers within each cluster and reflect on how the design knowledge is typically achieved and reported in these clusters of papers, but also how the communication of their research contributions could be further improved for practitioners and researchers to 
 utilize and build on.

In the remainder of this paper, we first present background on design science and our conceptualization of it by means of a visual abstract template (Section~\ref{sec:background}). 
We then describe our methodology for generating visual abstracts for the cohort of ACM distinguished papers we studied (Section~\ref{sec:method}), and 
use the information highlighted by the abstracts to extract the design knowledge in each paper. Finally, we cluster the papers by the design knowledge produced (Section~\ref{sec:result}).
We interpret and discuss the implications of our findings (Section~\ref{sec:discussion} and~\ref{sec:recommendations}), outline the limitations of our study (Section~\ref{sec:limitations}), and discuss related work (Section~\ref{sec:related_work}) before concluding the paper (Section~\ref{sec:conclusion}).


\section{Background}
\label{sec:background}
 
Our conceptualization of design science in software engineering, which our analysis is based on, was formed from a thorough review of the literature and a series of internal group workshops on the topic. This work helped us develop a visual abstract template to use as a lens for communicating and assessing research contributions~\cite{StoreyESEM17}. In this section, we summarize the findings by giving a brief introduction to design science and the visual abstract template. We use the term \emph{design knowledge} to refer to the knowledge produced in design science research.

\subsection{Design science}
\label{sec:design_science}

\begin{figure*}
\centering
  \includegraphics[width=0.65\textwidth]{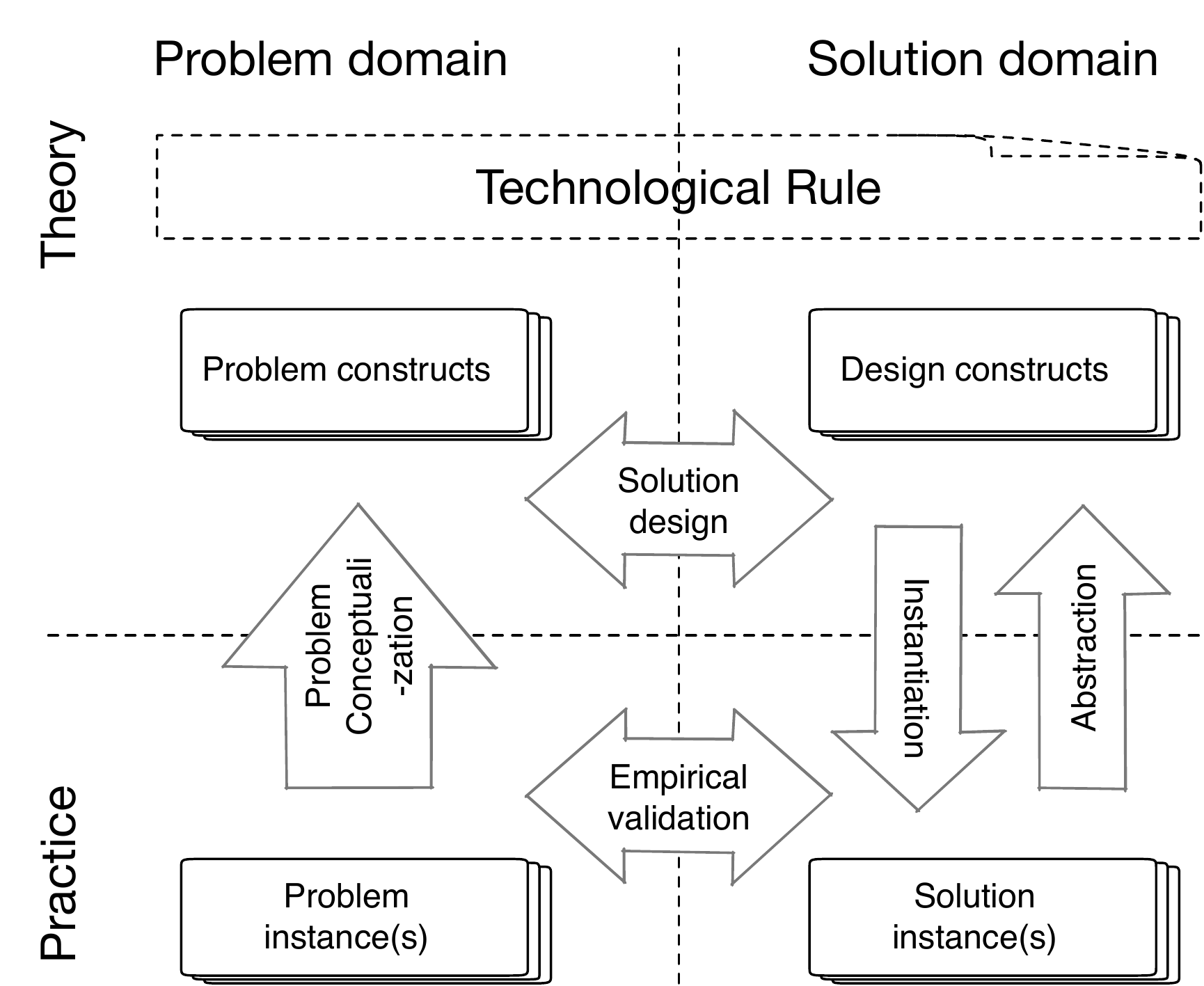}

\caption{An illustration of the interplay between problem and solution as well as between theory and practice in design science research. The arrows illustrate the knowledge-creating activities, and the boxes represent the levels and types of knowledge that is created.}
\label{fig:contributiontypes}       

\end{figure*}

The mission of design science is to solve real-world problems.  Hence, design science researchers aim to develop general design knowledge in a specific field to help practitioners create solutions to their problems. In Figure~\ref{fig:contributiontypes}, we illustrate the relationship between the \emph{problem domain} and \emph{solution domain}, as well as between \emph{theory} and \emph{practice}. The arrows in the figure represent different types of contributions of design science research, i.e., problem conceptualization,  solution design, instantiation, abstraction, and validation. 

Design knowledge is holistic and heuristic by its nature, and must be justified by in-context validations~\cite{wieringa_what_2014,aken_management_2004}. The term holistic is used by van Aken~\cite{aken_management_2004} and refers to the ``magic'' aspect of design knowledge, implying that we never fully understand why a certain solution works in a specific context. 
There will always be hidden context factors that affect a problem-solution pair~\cite{dyba_what_2012}. As a consequence, we can never prove the effect of a solution conclusively, and must rely on heuristic prescriptions. By evaluating multiple problem-solution pairs matching a given prescription, our understanding about that prescription increases. 
Design knowledge can be expressed in terms of \emph{technological rules}~\cite{aken_management_2004}, which are rules that capture general knowledge about the \emph{mappings} between \emph{problems} and proposed \emph{solutions}.

Van Aken describes the typical design science strategy to be the multiple case study~\cite{aken_management_2004}, which can be compared with alpha and beta testing in clinical research, i.e., first case and succeeding cases. Rather than proving theory, design science research strives to refine theory, i.e., finding answers to questions about why, when, and where a solution may or may not work.  Each new case adds insights that can refine the technological rule until saturation is achieved~\cite{aken_management_2004}. Gregor and Hevner present a similar view of knowledge growth through multiple design cycles~\cite{gregor_positioning_2013}. Wieringa~\cite{wieringa_technical_2012} and Johannesson~\cite{johannesson_introduction_2014} discuss action research as one of several empirical methodologies that can be used to produce design knowledge. Sein~\emph{et al.} \cite{Sein2011} propose how design science can be adapted by action research to emphasise the construction of artefacts in design science. 
However, action research does not explicitly aim to develop knowledge that can be transferred to other contexts, but rather it tries to make a change in one specific local context. 

\subsection{The design science visual abstract template}
\label{sec:VA}

\begin{figure*}
  \includegraphics[width=1.0\textwidth]{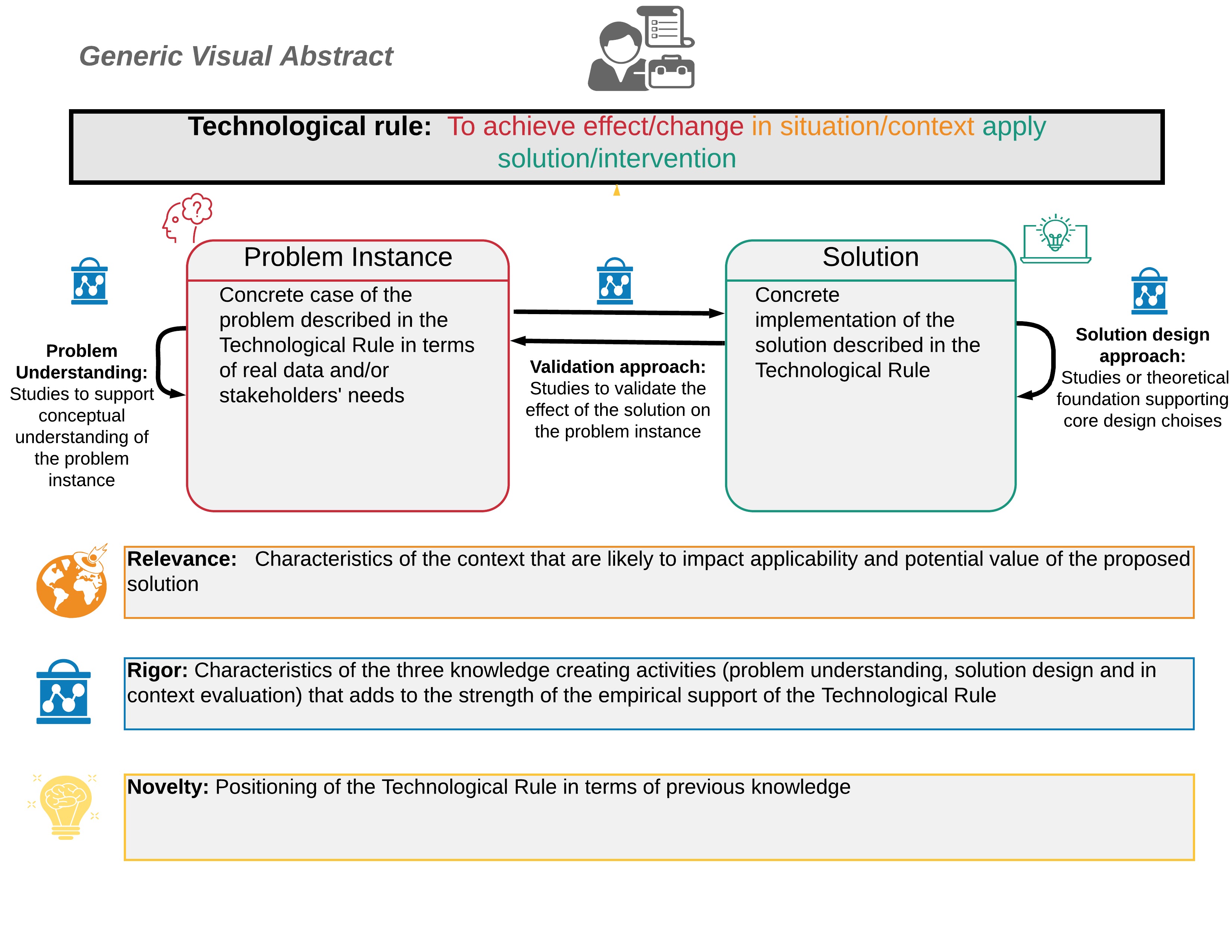}
\caption{The visual abstract template~\cite{StoreyESEM17}
capturing 1) the theory proposed or refined in terms of a technological rule; 2) the empirical contribution of the study in terms of a problem-solution instance and the corresponding design and validation cycles; and
3) support for the assessment of the value of the produced knowledge in terms of relevance, rigor, and novelty.
}
\label{fig:VA-template}      
\end{figure*}

The visual abstract template, shown in Figure~\ref{fig:VA-template}, captures three main aspects of design science contributions: 1) the theory proposed or refined in terms of a technological rule; 2) the empirical contribution of the study in terms of 
one or more instances of a problem-solution pair and the corresponding design and validation cycles; and
3) support for the assessment of the value of the produced knowledge in terms of relevance, rigor, and novelty. 
While adhering to the design science paradigm puts the focus on how to \emph{produce} and \emph{assess} design knowledge (i.e., technological rules), our visual abstract template is designed to help researchers effectively \emph{communicate} as well as justify design knowledge.
It also helps highlight which instantiations of the rule have been studied and how they were validated, how problem understanding was achieved, and what foundations for the proposed solution were considered. 
In the visual abstract template, the researcher is encouraged to reflect on how a study adds new knowledge to the general theory (i.e., the constructs of the technological rule) and to be aware of the relationship between the general rule and its instantiation (the studied problem-solution pair).

\subsubsection{The technological rule}
In line with van Aken~\cite{aken_management_2004}, our visual abstract template emphasizes technological rules (the top box in Figure~\ref{fig:VA-template}) as the main takeaway of design science within software engineering research. A technological rule can be expressed in the form:  
\emph{To achieve \textless Effect \textgreater ~ in \textless Situation \textgreater~apply \textless Intervention\textgreater}. 
Here, a class of software engineering problems is generalized to a stakeholder's desired effect of applying a potential intervention in a specified situation. 
Making this problem generalization explicit helps the researcher identify and communicate the different value-creating aspects of a research study or program. 
Refinements or evaluation of the technological rule may be derived from any one of the three processes of \emph{problem understanding}, \emph{solution design}, or \emph{solution validation}, applied in each instantiation. 

Technological rules can be expressed at any convenient abstraction level and are hierarchically related to each other. However, technological rules expressed at a very high abstraction level (e.g., ``to produce software of high quality, apply good software engineering practices'') tend to be either too high-level or too bold (easy to debunk). 
On the other hand, the lower the abstraction level is the more narrow the scope is, and thus there is a risk that detailed rules lack relevance for most software engineers. 
Thus, it is important to explicitly formulate the technological rule when presenting design science research and to be consistent with it both when arguing for its relevance and novelty, as well as when presenting the empirical (or analytical) support for the claims. 

\subsubsection{The problem-solution pair}

The main body of the visual abstract template (the middle section in Figure~\ref{fig:VA-template}) focuses on the empirical contribution of one or more studies, and is composed of two boxes for the problem-solution instantiation of the technological rule and three corresponding descriptions of the knowledge-creating activities, problem understanding, solution design, and validation.
 
\subsubsection{The assessment boxes}

The ultimate goal of design science research is to produce general design knowledge rather than to solve the problems of the unique instances. Thus, the value of the research should be assessed with respect to the technological rule (i.e., design knowledge) produced. The information in the three assessment boxes (the bottom of Figure~\ref{fig:VA-template}) aims to help the reader make an assessment that is relevant for their context. Hevner presents three research cycles in the conceptual model of design science, namely the \emph{relevance}, \emph{rigor}, and \emph{design} cycles~\cite{hevner_design_2004}. We propose that the contributions of design science research be assessed accordingly with respect to \emph{relevance, rigor}, and \emph{novelty}. 

The relevance box aims to support answering the question \emph{To whom is this technological rule relevant?} Relevance is a subjective concept and we are not
striving to find a general definition. Instead we suggest that the basic information needed to assess relevance of a research contribution is the potential effect of the proposed intervention combined with the addressed context factors. The relevance of a research contribution could be viewed from two perspectives: the targeted practitioner's perspective, and the research community's perspective. From the individual practitioner's point of view, the relevance of a research contribution is assessed by comparing their specific context with the one described in the research report. For the research community, a measure of relevance often relates to how common the studied problem is. To enable both types of assessment, relevant context factors need to be reported. A taxonomy of context factors in software engineering was proposed by Petersen and Wohlin~\cite{petersen_context_2009}. However, as discussed by Dyb\aa ~et al.~\cite{dyba_what_2012}, not all context factors are helpful in making this assessment; only those that are critical for either the applicability of the solution or for the potential effect of applying a solution should be reported. SERP is an initiative to guide the systematic development of more focused context taxonomies~\cite{petersen_finding_2014} and SERP taxonomies have been proposed in multiple subdomains of software engineering~\cite{engstrom_SERP-test_2017,ali_search_2019,rico_taxonomy_2019}. 

 The rigor box aims to support answering the question \emph{How mature is the technological rule?} Rigor of a design science study refers to the strength of the added support for the technological rule and may be assessed with respect to all of the three knowledge-creating activities: problem understanding, solution design, and solution validation. However, solution design is a creative process that does not necessarily add to the rigor of a study. One aspect of rigor in the design activity could be the extent to which the design is built on prior design knowledge. Also, the consideration of alternative solutions could be taken into account. On the other hand, the other two activities---problem understanding and solution validation---are based on common empirical methods on which relevant validity criteria (e.g., construct validity) can be applied. Note that the template only captures the claims made in the paper, and the validity of the claims are assumed to be assessed in the peer review process. 
  
 The novelty box aims to capture the positioning of the technological rule in terms of previous knowledge, and it supports answering the question \emph{Are there other comparable rules (similar, more precise, or more general rules) that should also be considered when designing a similar solution in another context?} Technological rules may be expressed at several abstraction levels; thus, it is always possible to identify a lower abstraction level where a research contribution may be novel, but doing so may be at the cost of general relevance. 
For example, a technological rule that expresses the efficiency of a technique in general may be made more specialized if it instead expresses the efficiency in one specific type of project that has been studied. Then the relevance is less general, and the novelty may be increased since it is the first investigation at that level of detail. Similarly, rigor is increased since the claims are less bold. 
 
 To optimize rigor, novelty, and relevance of reported research, the researcher should strive to express the technological rule at the highest useful abstraction level, i.e., a level at which it is novel, the provided evidence gives strong support and it is not debunked by previous studies (or common sense). However, adding empirical support for existing but under-evaluated technological rules has value, making novelty less important than the rigor and relevance criteria. To this extent, replication of experiments has been discussed~\cite{Carver2014,juristo2010replication,shull2008} and is encouraged\footnote{https://2018.fseconference.org/track/rosefest-2018} by the software engineering community. The incremental adding of empirical support for a technological rule could be referred to as conceptual replication in which the same research question is evaluated by using a different study design, as discussed by Shull \emph{et al.}~\cite{shull2008}.

\section{Methodology}
\label{sec:method}

The main goal of this paper was to investigate how well software engineering (SE) research contributions are aligned with the design science paradigm. 

As part of this work, we aimed to answer the following \textbf{research questions}: 
\begin{enumerate}[RQ1]	
\item From a design science perspective, what \emph{types of contributions} do we find in the SE community? 
\item In papers that present design knowledge, how clearly are the theoretical contributions (i.e., the \emph{technological rules}) defined in these papers?
\item How are \emph{novelty}, \emph{relevance} and \emph{rigor} discussed in papers with design knowledge contributions?
\end{enumerate}

\begin{figure*}
  \includegraphics[width=1\textwidth]{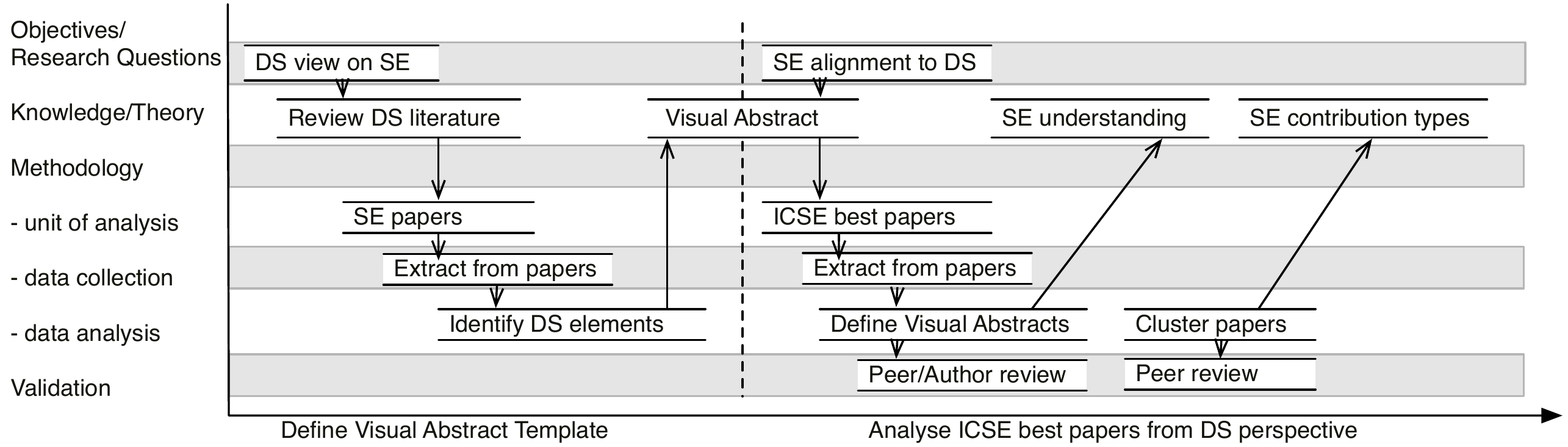}
\caption{The approach followed to develop the initial version of the visual abstract (the left side) and the main steps of the research presented in this paper (the right side).}
\label{fig:methodology}       
\end{figure*}

As mentioned above, our earlier research produced a visual abstract template for communicating design science research~\cite{StoreyESEM17}.
Earlier research and the research presented in this paper are illustrated in Figure~\ref{fig:methodology}. The left side of the figure shows the steps we followed to arrive at the initial version of the visual abstract template, while the right side shows the steps we followed in the research reported in this paper, which also helped us refine the visual abstract template and instructions for filling it out. 

We used this visual abstract template to describe the research contributions in
a particular set of papers from the main technical track at the ICSE conference: those that were selected as the top 10\% of papers (i.e., papers designated the ACM SIGSOFT Distinguished Paper Award\footnote{https://www.sigsoft.org/awards/distinguishedPaperAward.html}) across five years of the conference (2014--2018 inclusive)
We chose ICSE because it is considered to be one of the top publishing venues in software engineering that covers a broad set of diverse topics, and we chose the ``best'' of those papers because we expected this cohort would represent exemplars of fine research.
In total, we considered and applied the visual abstract template to describe the research contributions across 38 papers, which are listed separately at the end of the references for this paper.

The process for defining the visual abstracts was as follows. 
Each paper in the cohort of ICSE distinguished papers from 2014--2018 was randomly assigned to two reviewers (among the authors of this paper). As the work was not about judging papers, but about understanding, we did not care for any conflicts of interest. 
The two reviewers independently extracted information from the papers to answer the set of \emph{design science questions} listed in Table~\ref{tab:method-vaquestions}. 
This set of questions 
 map to the different components in the visual abstract template. Thus, we defined a visual abstract for each paper, which we iterated until we arrived at an agreement for a shared response to these questions, seeking additional input through reviews by the rest of our research team and expert opinions for papers on topics unfamiliar to us. 

The answers to the questions were captured in a spreadsheet to facilitate future analysis as well as ongoing review and internal auditing of our process.
Our combined responses were then used to populate the visual abstract template for each paper.
The collection of visual abstracts for all of the papers is available online at \url{dsse.org}, which constitutes our understanding of the analyzed software engineering research from a design science perspective.

As part of our analysis, we confirmed our interpretations of the 2014 ICSE distinguished papers with the original authors. We heard back from half of the authors of this set of papers, who confirmed the accuracy of our responses (mentioning minor improvements only). 
We assessed this feedback as a preliminary validation of our process and did not feel the need to repeat this step for the other papers---in each case, the abstracts for all papers we studied are available online\footnote{dsse.org} and the authors may comment publicly on our interpretations if they choose.

\begin{table}
\caption{Characterizing research through a design science lens: The answers to the following questions were used to populate a visual abstract for each paper. }
\label{tab:method-vaquestions}     
\begin{tabular}{p{.75cm}p{10.25cm}}
\hline\noalign{\smallskip}
1. & \emph{Problem instance} \\
1.1 & What problem is addressed in the paper? 
(Describe in terms of the concrete instance of the problem studied.) \\
\hline\noalign{\smallskip}
2. & \emph{Problem understanding approach} \\
2.1 & How did the authors gain an understanding of the problem? \\
\hline\noalign{\smallskip}
3. & \emph{Proposed solution(s)} \\
3.1 & What intervention(s) was proposed to solve the identified problem? \\
\hline\noalign{\smallskip}
4. & \emph{Design approach} \\
4.1 & How did the authors arrive at their proposed solution? \\
\hline\noalign{\smallskip}
5. & \emph{Validation approach} \\
5.1 & How did the authors apply the intervention/solution to the problem instance to validate it? \\
\hline\noalign{\smallskip}
6. & \emph{The Technological Rule} \\
6.1 & What effect do they wish to achieve through their research? \\
6.2 & In what situations does this rule apply? \\
6.3 & In summary, what is the proposed solution in the paper? \\
\hline\noalign{\smallskip}
7. & \emph{Relevance, convincing the target stakeholder} \\
7.1 & What class of problems and solutions are captured by the technological rule? \\
7.2 & To whom are those problem-solution pairs relevant? \\
7.3 & How do the authors convince their readers that the problem-solution pair is relevant to those stakeholders? \\
\hline\noalign{\smallskip}
8. & \emph{Rigor} \\
8.1 & What actions have been taken to ensure the understanding of the problem instance is valid?  \\
8.2 & What actions have been followed to ensure the intervention is a valid solution to the problem instance? \\
8.3 & What actions have been taken to validate the design choices? \\
\hline\noalign{\smallskip}
9. & \emph{Novelty} \\
9.1 & What are the novel contributions in the paper? \\
\hline\noalign{\smallskip}
\end{tabular}
\end{table}

%
%

Once we finished creating all the visual abstracts, we began clustering the papers (see the rightmost part of Figure~\ref{fig:methodology}). Note as we answered questions in Table~\ref{tab:method-vaquestions}, we presented our answers to other members in our research group for feedback, which in many cases helped us refine the responses.
We also printed the visual abstracts we created for each paper (in miniature), and 
working as a group in a face-to-face meeting, we sorted the visual abstracts into bundles to identify \emph{clusters} representing different types of design science contributions.

Following our face-to-face visual abstract clustering activity, we worked again in pairs to inspect each of the papers in the clusters in order to confirm whether we had categorized them correctly.
Again, we reviewed and confirmed the categorization of each paper as a group. 
 During this confirmation process, we refined our categorization and collapsed two categories into one: we combined papers that were initially classified as exploratory with papers that we initially thought were design science contributions in terms of problem understanding but on reflection were better framed through an explanatory lens as the investigated problems were not linked to a specific solution.  
We present the stable clusters that emerged from these activities in the following section of this paper.

\section{Results from the paper cluster analysis}
\label{sec:result}

 Overall we identified five clusters, described in detail below, based on our analysis of how each paper contributed to the extracted technological rule. Note the rules are not extracted by the original authors of the papers but by us for the purpose of this particular review.  

\begin{enumerate}
	\item \emph{Problem-solution pair}: this cluster represents papers that equally balance their focus on problem instance and solution. 
	\item \emph{Solution validation}: this cluster is characterized by papers that concentrate largely on the solution and its validation, rather than on problem understanding.	
	\item \emph{Solution design}: papers in this cluster focus on the design of the solution rather than on problem understanding or solution validation.	
	\item \emph{Descriptive}: these papers address a general software engineering phenomenon rather than a specific instance of a problem-solution pair. 	
	\item \emph{Meta}: this cluster of papers may be any of the types above but are aimed at research insights for researchers rather than for practitioners. 
\end{enumerate}

\begin{figure*}
\centering
  \includegraphics[width=0.65\textwidth]{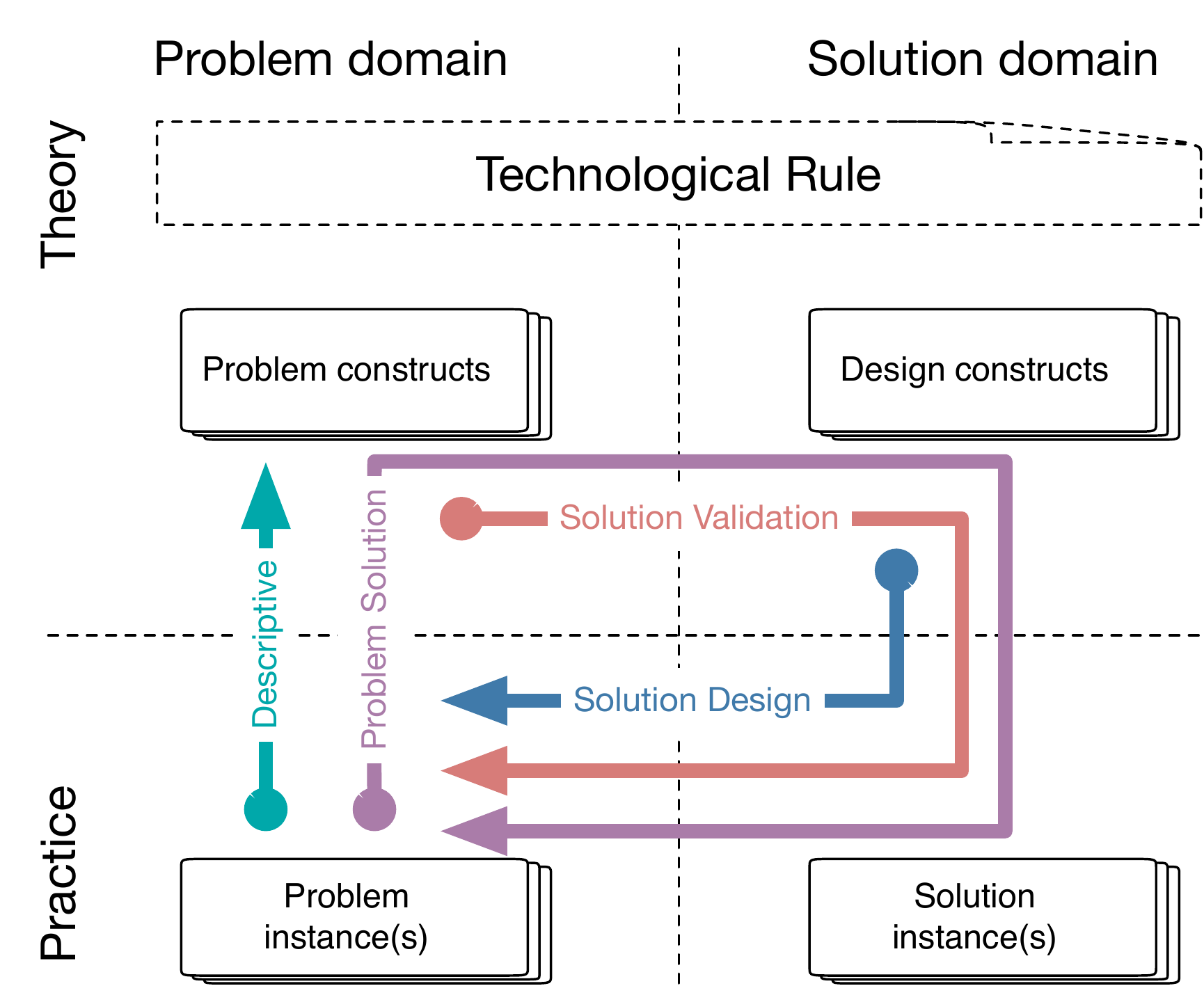}

\caption{An illustration of how the identified clusters map to the problem/solution and the practice/theory axes respectively. The arrows show how typical studies in each cluster traverse the four quadrants (1.~practical problem, 2.~conceptual problem description, 3.~general solution design, and 4.~instantiated solution).}
\label{fig:clusterpath}       
\end{figure*}

Figure~\ref{fig:clusterpath} illustrates how the first four clusters (1--4) map to the design science view, including both the problem-solution dimension and the general-specific one. Clusters 1--3 all represent different types of design science research since the papers in these clusters consider explicit problem-solution pairs. Papers in the fourth cluster provide explanatory knowledge and, although such knowledge may support software engineering solution design, they are better framed through an explanatory lens.
Cluster 5 is not represented in this figure as these papers produce knowledge on software engineering research rather than on software engineering practice. 

\begin{figure*}
  \includegraphics[width=1\textwidth]{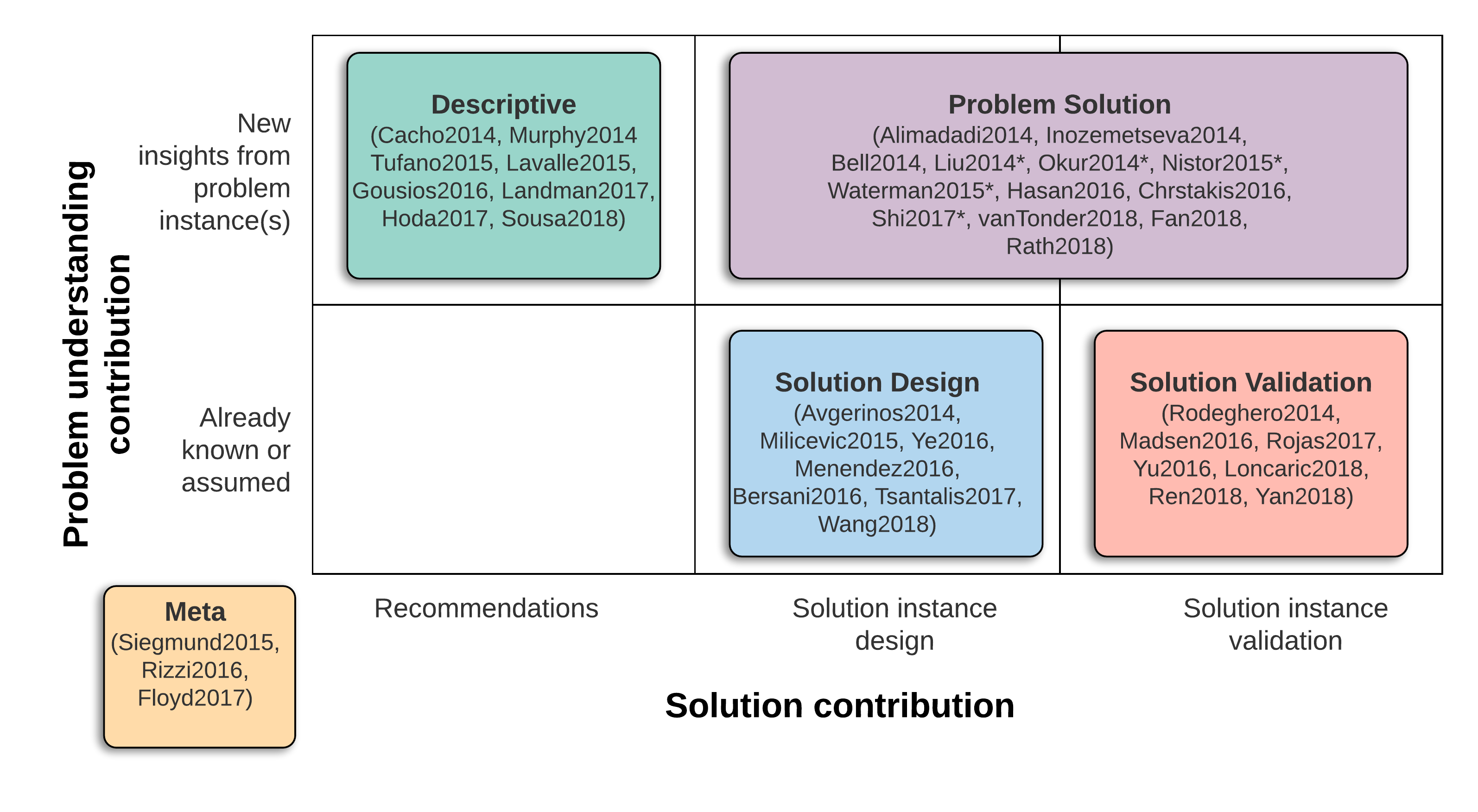}
  \vspace{-30pt}
\caption{The main clusters that emerged from our analysis of the papers, showing the key design science contributions in terms of problem understanding insights and solution recommendations, design and/or validation.}
\label{fig:paper-clusters}       
\end{figure*}

Figure~\ref{fig:paper-clusters} shows a visual representation of the main clusters that emerged from our analysis along with a listing of which papers (first author/year) belong to the different clusters. 
The two axes of this graph are defined as follows: the x-axis captures the solution contribution ranging from high-level recommendations, to more concrete solutions that are designed and may be validated;  and the y-axis indicates the problem understanding contribution whereby the problem is already known or assumed, to where new insights are produced from the research.  
 
A more detailed and nuanced description for each cluster is provided below. For each cluster we refer to examples of papers and include one visual abstract to showcase the design knowledge that is or is not captured by each cluster. Note that the example visual abstracts show what was reported in the papers and thus do not perfectly match the intention of all elements in the visual abstract template (see Figure~\ref{fig:VA-template}). 

\subsection{Problem-solution pair}
\label{sec:problem_solution}

\begin{figure*}
\centering
	\includegraphics[width=01.0\textwidth]{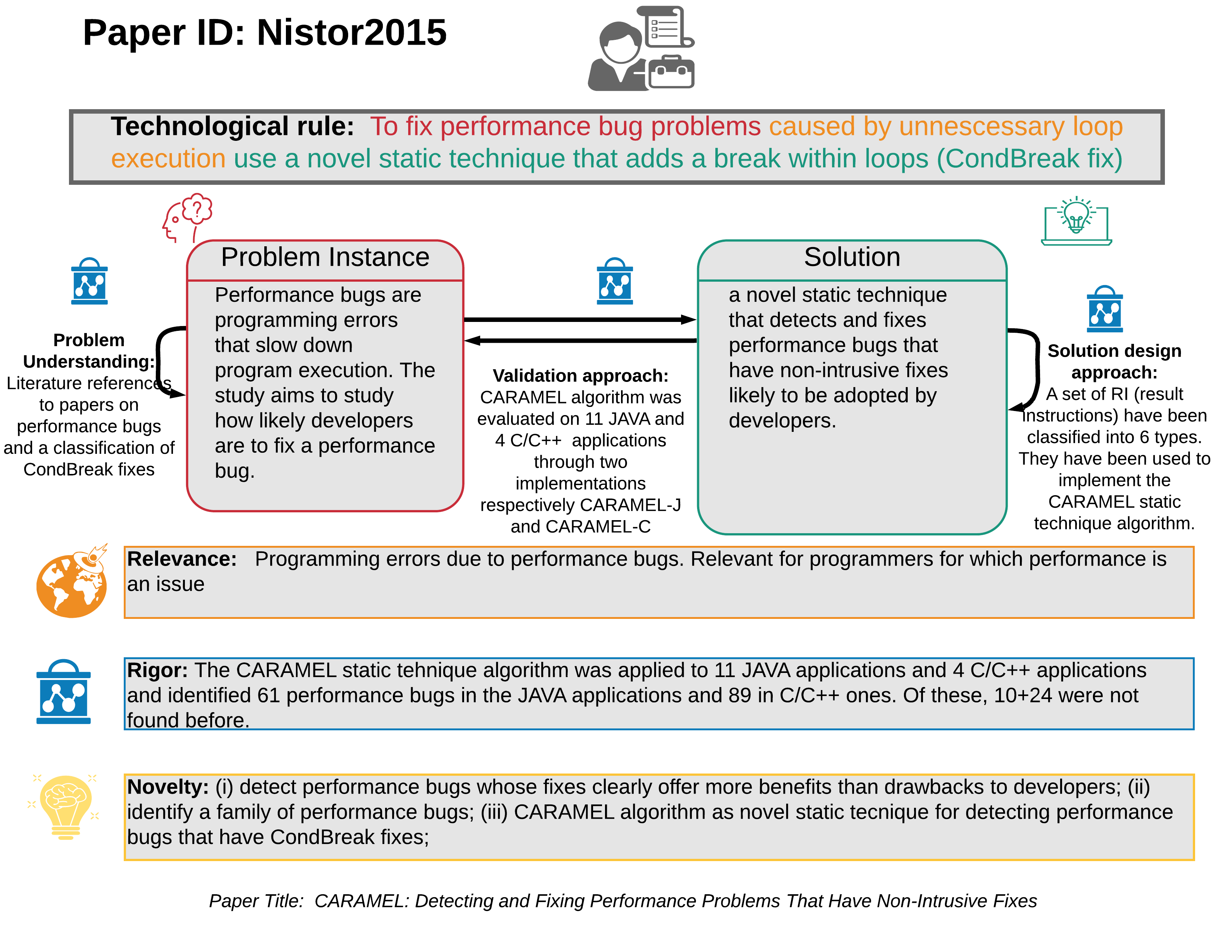}
	\caption{Visual abstract of a typical paper in the problem solution cluster, Nistor \emph{et al.}~\protect\citeI{Nistor2015}}
	\label{fig:VA-problemSolution}  
\end{figure*}

For the papers in this cluster, a problem instance is identified and investigated to gain a generalized problem formulation matching the proposed solution.  A solution is proposed, designed and implemented, then validated rigorously in-context through empirical methods. It is the most populated cluster, indicating that many software engineering papers can be framed in accordance with the design science paradigm.

The technological rule is defined quite clearly in all of the papers belonging to this cluster and is in most cases a new proposal of either a tool or methodological approach to adopt to solve the problem instance (see Figure~\ref{fig:VA-problemSolution}). Consequently, the relation among problem (e.g., performance bug problems) and solution (e.g., novel static analysis technique \textsc{caramel} that detects and fixes performance bugs) is explicit. 

Solutions are geared towards both practitioners and researchers, making it explicit and easy for a stakeholder to assess the relevance of the rule for their specific case. The solutions are mainly validated by conducting case studies on real projects \citeI{Nistor2015} or controlled experiments \citeI{Alimadadi2014,Bell2014}.

In some cases alternative solutions are compared to the proposals made. For example, Rath \emph{et al.} \citeI{Rath2018} considered alternative information retrieval techniques and classifiers during the design of their solution, and used precision/recall values collected from all the compared solutions to develop their classifier. 

A representative example for this cluster is the paper by Nistor \emph{et al.}~\citeI{Nistor2015}. Given the problem instance, where analysis of the related literature points out that performance bugs are programming errors that slow down program execution, the authors investigate how likely developers are to fix a performance bug. The solution proposed is a novel static analysis technique, \textsc{caramel}, which is able to detect and fix performance bugs. Nistor \emph{et al.}  designed a set of case studies to validate the tool on a set of Java and C++ applications. The visual abstract is shown in Figure \ref{fig:VA-problemSolution}. Other visual abstracts in this cluster (and other clusters) are available on our online website\footnote{dsse.org}.

In summary, the problem solution cluster papers can be seen as presenting complete design science contributions, considering both the general and specific aspects of a problem-solution pair investigated in context, with implications for  researchers and practitioners.
 
\subsection{Solution validation} 
\label{sec:solution_validation}

Papers in the solution validation cluster mainly focus on refining a previously proposed (in many cases implicit) technological rule. The problem is implicitly derived from a previous solution and its limitations, rather than from an observed problem instance. Accordingly, in most cases, the problem is motivated by a general statement at an abstract level, making claims about ``many bugs...'' or ``it is hard to...''. Some of the papers underpin these claims with references to empirical studies, either the authors' own studies, or from the literature, while others ground the motivation in what is assumed to be generally ``known''. 

\begin{figure*}
  \includegraphics[width=01.0\textwidth]{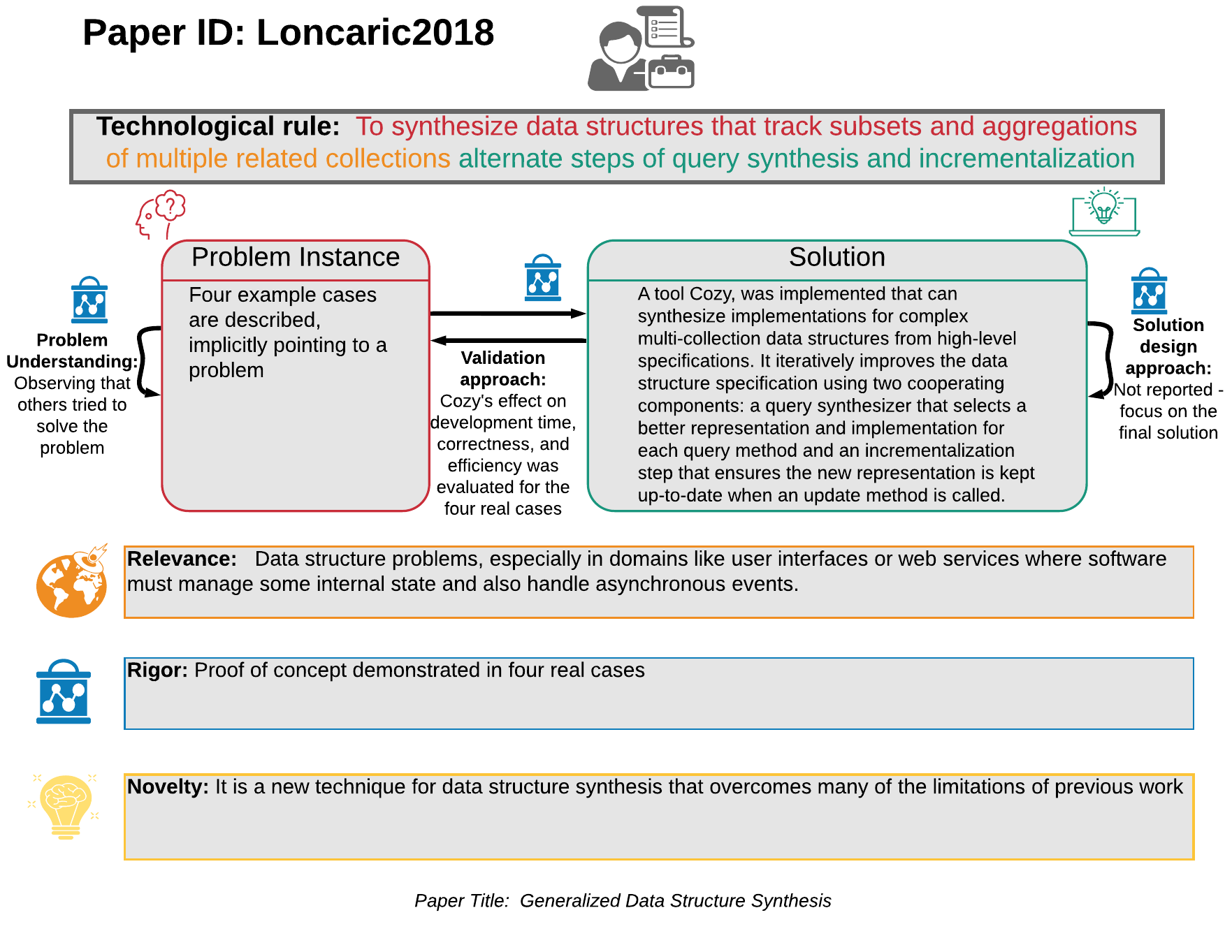}
\caption{Visual abstract of a typical paper in the cluster of solution validation studies, Loncaric \emph{et al.}~\protect\citeI{Loncaric2018} }
\label{fig:VA-Loncaric2018}       
\end{figure*}

As a typical example, Loncaric \emph{et al.}~\citeI{Loncaric2018}, identify that others have tried to automate the synthesis of data structures, and present a  tool that embeds a new technique that overcomes the limitations of previous work. A proof of concept is demonstrated in four real cases. The corresponding visual abstract is presented in Figure \ref{fig:VA-Loncaric2018}.

Note that some papers in this cluster focus on understanding the problem with previous solutions, with the aim to improve the solution or come up with a new one. For example, Rodeghero \emph{et al.}~\citeI{Rodeghero2014} attempt to improve code summarization techniques for program comprehension. They perform an extensive eye-tracking study to design a code summarization tool.  

The technological rules are mostly implicit in these papers. As they are related to problems with existing solutions, rather than original problems in the SE domain, the presentation of the solutions are mostly related to previous solutions. A technological rule can sometimes be derived indirectly, through the aim of an earlier solution, but it is rarely defined explicitly. 

The papers in this cluster discuss relevance to research explicitly, while the relevance to practice is mostly discussed indirectly, and at a high abstraction level. For example, Rojas \emph{et al.} ~\citeI{Rojas2017} claim that writing good test cases and generating mutations is hard and boring, and thus they propose a gaming approach to make this more enjoyable and better. The validation is conducted, testing a specific code instance, while the original problem is rooted in high-level common sense knowledge. However, there are other papers in the cluster that back up the problem through evidence, such as a vulnerability database, used by Yan \emph{et al.}~\citeI{Yan2018} to motivate addressing the vulnerability problem of Use-After-Free pointers.

In summary, the solution validation papers focus on refining an existing technological rule. The motivating problem is mostly expressed in terms of high-level knowledge, rather than specific instances, although some papers refer to empirical evidence for the existence and relevance of the problem. The more specific problem description is often related to problems with previous solutions. The papers clearly show a design science character, although they are at risk of solving academic problems, rather than practitioners' problem instances.

\subsection{Solution design}
\label{sec:solution_design}

\begin{figure*}
	\includegraphics[width=01.0\textwidth]{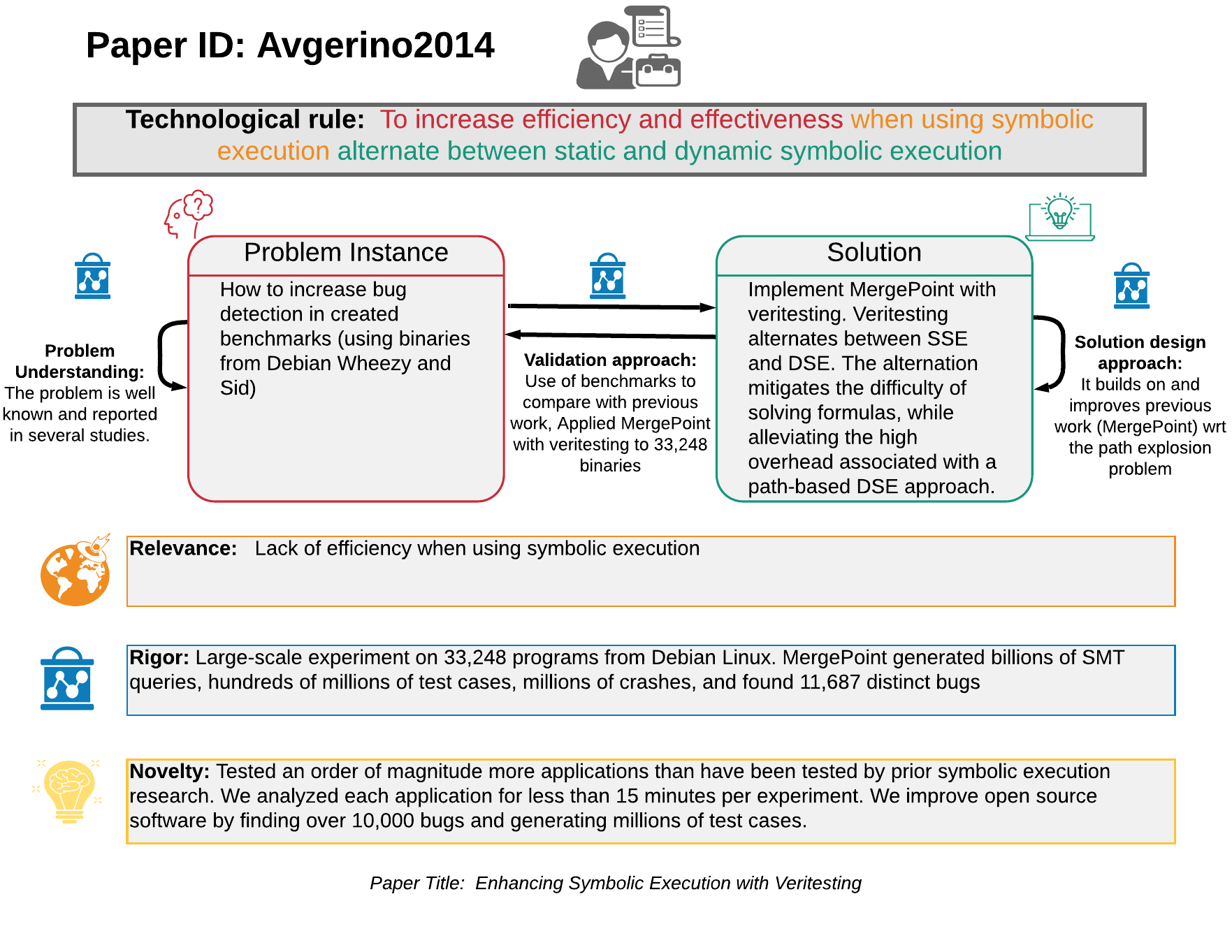}
	\caption{Visual abstract of a typical paper in the solution design cluster, Avgerinos~\emph{et al.} \protect\citeI{Avgerinos2014}}
	\label{fig:VA-solutiondesign} 
\end{figure*}

The papers in this cluster present details of a new instantiation of a general solution. For example, Avgerinos \emph{et al.}~\citeI{Avgerinos2014} present a new way of testing with symbolic execution, see Figure~\ref{fig:VA-solutiondesign}. 
The presented approach finds more bugs than the previously available methods. 
However,  the need for this tool was not explicitly stated and the authors perhaps assume the need is clear.

Similarly, in Bersani \emph{et al.}~\citeI{Bersani2016}, these authors propose a new semantics for metric temporal logic (MTL) called Lazy Semantics for addressing memory scalability. The proposal builds on previous research and is focused on the solution, i.e., a new trace checking algorithm.
A similar observation can be made for  analysis and validation, i.e., the analysis in Avgerinos \emph{et al.}~\citeI{Avgerinos2014} is conducted by using the proposed solution on a rather large code base and using well known metrics such as number of faults found, node coverage, and path coverage. Whereas in Bersani \emph{et al.}~\citeI{Bersani2016}, the validation is carried out comparing the designed solution with other, point-based semantics. 

For papers in this clusters, the problem is not explicitly formulated, but it is more generally discussed in terms of, for example, decreasing the number of faults. The papers tend to describe the designed solutions in rather technical terms. This is also how the novelty typically is highlighted. 
Validations are typically conducted by applying the proposed solution on a code base and analyzing metrics of, e.g., the number of faults found in testing, and no humans are directly involved as subjects in validations. Empirical data for the validations are either obtained by technically measuring e.g.\ execution time, or by using data already published in programmer-forums.

In summary, the solution design papers focus on low level technological rules. The motivating problem is in most cases technical details of a solution to a more general problem. While the validity of the general solution is implicit, the low level solution is often validated through controlled experiments or benchmarking in a laboratory setting.
The papers clearly show a design science character, although at a low abstraction level, inducing a risk of losing the big picture in favor of fine-tuning.  

\subsection{Descriptive}
\label{sec:descriptive}

\begin{figure*}
  \includegraphics[width=01.0\textwidth]{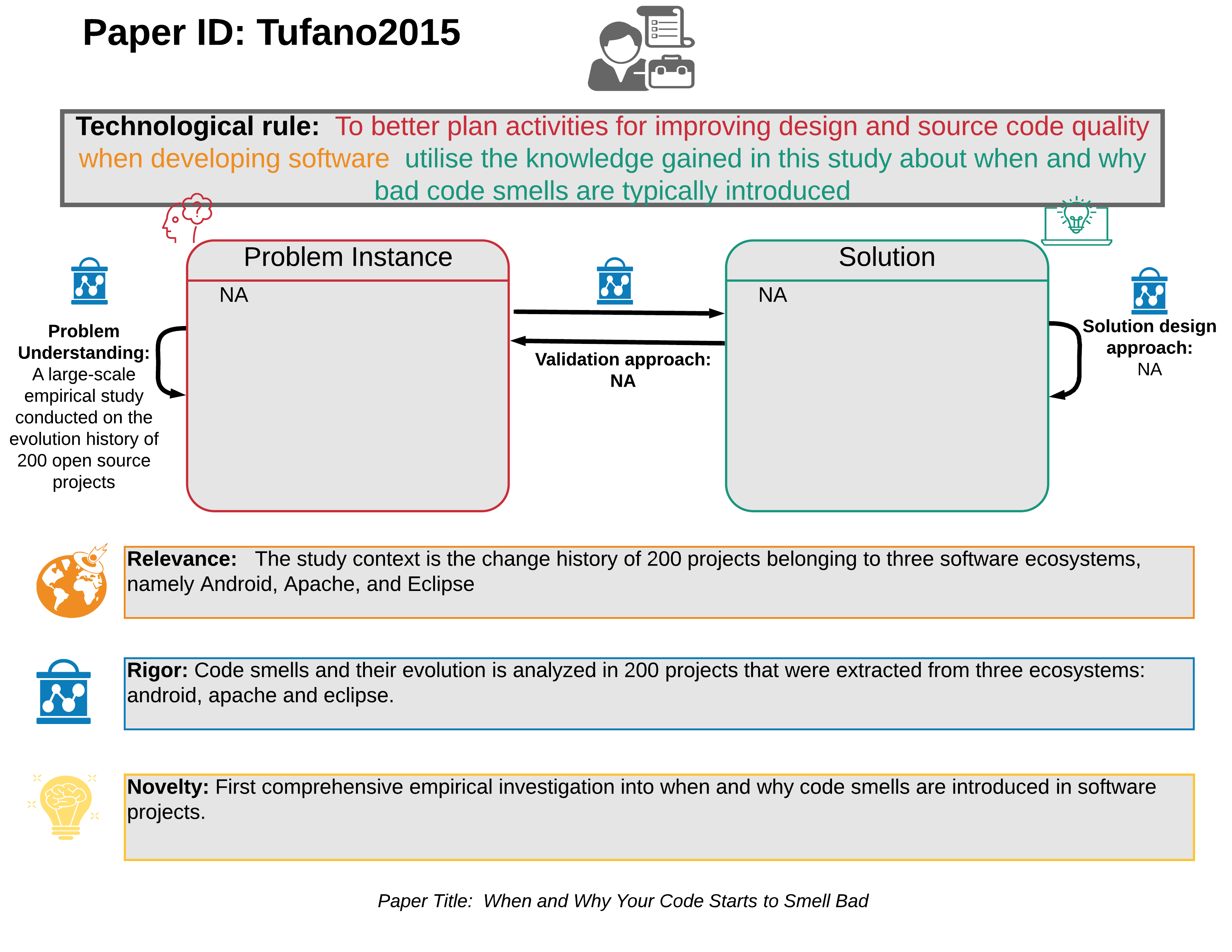}
\caption{Visual abstract of a typical paper in the cluster of descriptive studies, Tufano \emph{et al.}~\protect\citeI{Tufano2015}}
\label{fig:VA-problem} 
\end{figure*}

The papers categorized in this cluster develop an understanding of a software engineering phenomenon that is currently not well understood. Such research studies may expose problems that need to be addressed, or they may reveal practices or tools that could benefit other challenging software engineering scenarios. 

For example, Murphy \emph{et al.}~\citeI{Murphy-Hill2014} conducted a study of game developers and identify a number of recommendations for how game developers could be better supported through improved tools or practices, while Hoda \emph{et al.}~\citeI{Hoda2017} carried out a grounded theory study to achieve an understanding of how teams transition to agile.  

Concrete instances of software engineering phenomena have been studied in various ways. Gousios \emph{et al.}\citeI{Gousios2016} surveyed 4000 open source contributors to understand the pull-based code contribution process, Tufano \emph{et al.}~\citeI{Tufano2015} analyzed git commits from 200 open source repositories to investigate more about code smells, Cacho \emph{et al.}~\citeI{Cacho2014} studied changes to 119 versions of code extracted from 16 different projects to understand trade-offs between robustness and maintenance and Lavallee \emph{et al.}~\citeI{Lavallee2015} reported on a 10 month observational study of one software development team to understand why ``good developers write bad code''.

Figure~\ref{fig:VA-problem} shows a typical example of a visual abstract from this cluster. The theoretical contributions of these studies are explanatory problem characterizations. In four papers out of eight, a list of recommendations is provided as well. 
Thus, it is in most cases possible to derive several technological rules from one paper.
However, these technological rules are not instantiated or evaluated further, and neither are they highlighted as the main contributions of the reported studies. 

All papers in this cluster discuss relevance to practice: many explicitly discuss how common the phenomenon under study is (e.g., Gousios \emph{et al.}~\citeI{Gousios2016} show a diagram of the monthly growth of pull request usage on GitHub). Others implicitly highlight a knowledge gap assumed to be of importance (e.g., Lavallee \emph{et al.}~\citeI{Lavallee2015} pinpoint the lack of knowledge about the impact of organizational factors on software quality). 
Novelty or positioning is, on the other hand, not described in terms of the problem or the solution but about aspects of the study as a whole. 
Gousios \emph{et al.}~\citeI{Gousios2016}  add a \emph{novel perspective}, the contributors' code review,  Lavallée \emph{et al.}~ \citeI{Lavallee2015} add \emph{more empirical data} about organizational factors and software quality, and  Tufano \emph{et al.}~\citeI{Tufano2015}  claim to report the \emph{first empirical investigation} of how code smells evolve over time.

In summary, although the descriptive papers may contribute to design knowledge, i.e., understanding of conceptual problems and initial recommendations, design knowledge in the form of technological rules are not directly described in the papers. 
The main contributions are discussed in more general terms such as descriptions of the phenomenon under study (defined in the titles) and general information about the study approach and the studied instances (which often appears in the abstracts). Potential problems and their solutions are described in the discussion sections of the papers.
Their relevance to practice is in terms of the real-world problems or recommendations that other applications have that tend to be exposed by these kind of papers. Thus, such papers are typically reporting on exploratory research that may be quite high in novelty.

\subsection{Meta}
\label{sec:meta}

\begin{figure}[t]
  \centering
  \includegraphics[width=\columnwidth]{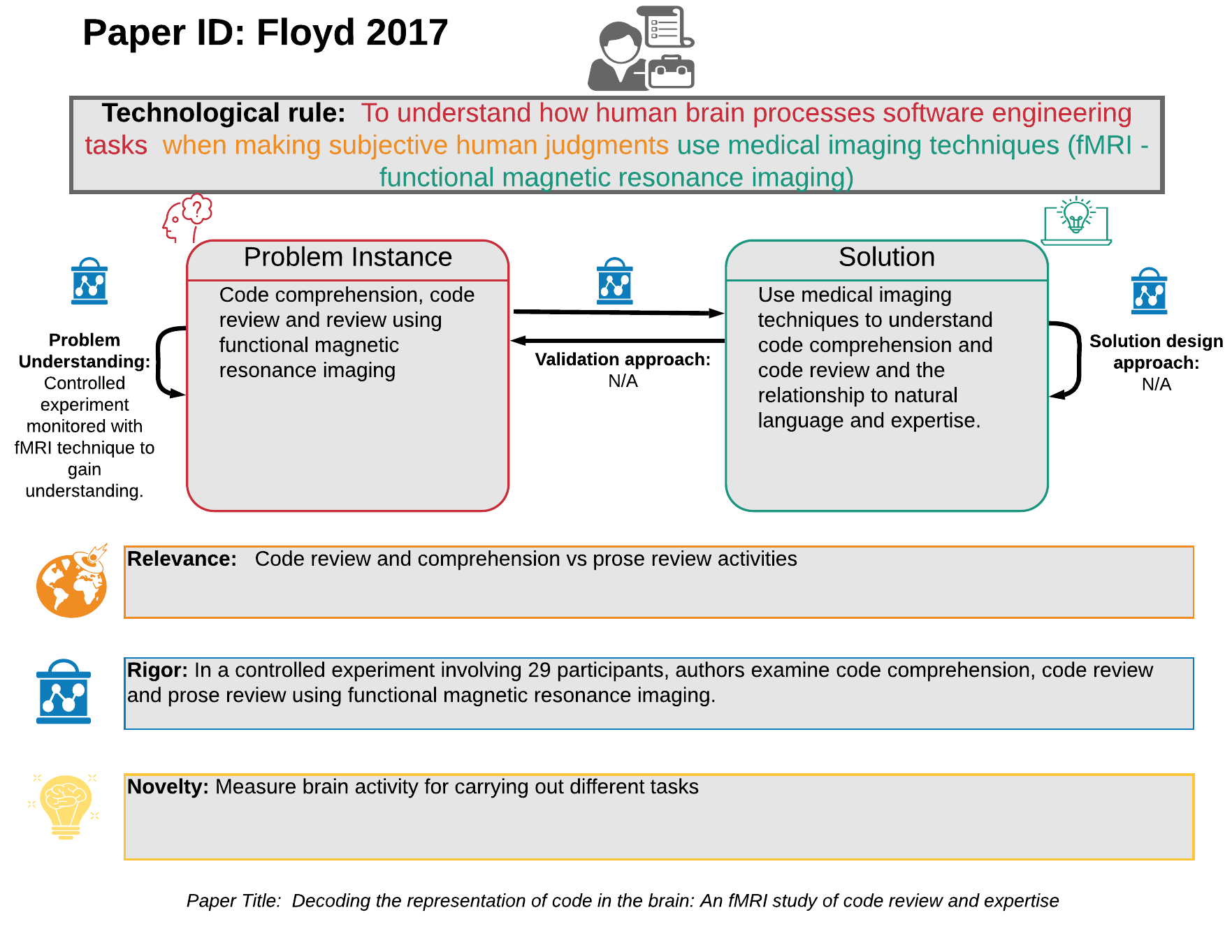}
  \caption{A typical example of a visual abstract in the Meta cluster, Floyd~\emph{et al.} \protect\citeI{Floyd2017}}
  \label{fig:meta-VA}
  \vspace{-8pt}
\end{figure}

Three papers in the set do not aim to identify or solve software engineering problems in the real world. Rather, these studies aim at identifying or solving problems which software engineering \emph{researchers} may experience. We therefore refer to them as Meta studies, i.e., addressing the meta level of software engineering \emph{research} in contrast to the primary level of software engineering \emph{practice}. 
Siegmund \emph{et al.}~\citeI{Siegmund2015} conducted a study that reveals how the software engineering research community lacks a consensus on internal and external validity.
Rizzi \emph{et al.}~\citeI{Rizzi2016} advise researchers how to improve the efficiency of tools that support large-scale trace checking.
Finally, Floyd \emph{et al.}~\citeI{Floyd2017} propose how fMRI methods can help software engineering gain more insights on how developers comprehend code, and in turn may improve comprehension activities. We show the visual abstract for the Floyd \emph{et al.}~\citeI{Floyd2017} paper in Figure~\ref{fig:meta-VA}.  

These papers address software engineering \emph{research} problems, and propose solutions for software engineering research. The design knowledge gained in these studies is primarily about the design of software engineering \emph{research}, and the key stakeholders of the technological rule is rather the \emph{researchers} than software engineers. Still, they fall under the design science paradigm and the Meta category of papers may show relevance to industry but in an indirect manner.

In summary, papers that we describe as Meta may fall under the design science research paradigm, leading to a technological rule with \emph{researchers} rather than software engineers as the key stakeholders.  


\section{Discussion: Design Science Contributions in Software Engineering}
\label{sec:discussion}

The long term goal of much software engineering research is to address real-world problems and provide useful recommendations on how to address those problems with evidence for benefits and potential weaknesses of those recommendations. 
Our analysis of ICSE distinguished papers reveals empirical contributions (RQ1) related to \emph{problem understanding}, \emph{solution design}, \emph{solution implementation}, and \emph{solution validation} (see path traversal in Figure~\ref{fig:clusterpath}). 
In some papers, all four activities are explored in equal depth while others focus on one or two activities, as shown in the clusterings above in Section \ref{sec:result}. All of those activities generate knowledge corresponding to the elements of our visual abstract template (see Figure~\ref{fig:VA-template}).
However, none of the papers are \emph{presented} in terms of these elements and we had to spend significant effort, using the questions in Table~\ref{tab:method-vaquestions}, to extract this knowledge in a systematic way. 
Extracting technological rules for the papers was also mostly quite challenging.
That said, applying the design science lens helped us notice and distinguish the different kinds of design contributions from the papers we analyzed, and guided our assessment of the papers in terms of research relevance, rigor and novelty. 
We discuss our experiences using the design science lens below and also the challenges we faced applying it to different types of papers. We also allude to experiences we have faced as researchers and reviewers of research papers below.

\subsection{Problem understanding and descriptive research in software engineering}

We found the design science paradigm helped us distinguish descriptive research contributions from prescriptive research contributions in the papers we analyzed.
Indeed eight of the papers we analyzed focused primarily on the understanding of software engineering problems or phenomenon that were not currently well understood. 
Descriptive papers are often labeled by the authors as ``exploratory'' research.
Often these papers do not only describe or expose specific problems or phenomenon, but they may also describe why or how certain solutions or interventions are used, and conclude with directions for future research or with recommendations for practitioners to consider (e.g., to use a studied intervention in a new context).

We struggled at first to apply the design science lens to some of these descriptive papers, as for most of them, no explicit intervention or recommendations were described or given. Articulating clear technological rules was not possible, as this research does not aim at producing design prescriptions (yet). 
However, on reflection we recognized that the design science lens helped us to recognize and appreciate the longer term goals behind this exploratory research that would later culminate in design knowledge. Sometimes we have found that descriptive research is under appreciated over prescriptive solutions, but understanding problems clearly is also an important research contribution in a field like software engineering that changes rapidly. In fact, often researchers are ``catching up'' to what is happening in industry and to recognize new emerging problems that may arise in industrial settings as tools and practices evolve.

Another cluster of papers we identified, 13 problem-solution pair papers, also contribute insights on problems experienced in software engineering projects. 
Many of the problem-solution papers derive problem insights from specific problem instances.  This was the biggest cluster of papers. 
The design science lens helped us to recognize and appreciate that these papers had contributions not just on the solution design and validation side, but also contributed or confirmed insights on a studied problem.
We have all had experiences when reviewing papers where a co-reviewer failed to recognize problem understanding contributions and argued that a given solution was either too trivial or poorly evaluated.   
As papers are written (and then typically read) in a linear fashion, losing track of the various contributions can happen.  For us, laying out the contributions visually (and by answering the questions we explicitly posed in Table 1) helped us to keep track of and appreciate contributions on both the problem and solution aspects.

\subsection{Solution design contributions in software engineering research}

The other two main clusters of papers that are aimed at improving software engineering practice are the 7 solution-design papers and 7 solution-validation papers. 
These papers contribute \emph{design knowledge} concerning an intervention and mostly rely on either previous research or accepted wisdom that the problem they address is in need of solving. 
For these clusters, the first questions in Table 1 about the problem instance addressed and problem understanding approach did not always have an explicit answer in the paper. However, to conduct an empirical validation of the design, some kind of instantiation of the problem is required and we referred these instances when extracting information about problem instance and problem understanding for our analysis. We found this to be an effective way to address distances between the abstraction level of the proposed technological rule and its empirical validation.
Papers without specified problem instances are at risk of proposing solutions, which do not respond to real software engineering problems.

\subsection{Identifying technological rules from software engineering research}
For most papers, we were able to extract \emph{technological rules} from the presented research. However, none of the papers had any conclusion or recommendation in such a condensed form (see RQ2). In some cases, the abstracts and introduction sections were written clearly enough that we could identify the intended effect, the situation and the proposed solution intervention presented in the paper. Moreover, when research goals and questions were explicitly stated, technological rules were easier to formulate. 
Other papers required more detailed reading to extract the needed information. 
In some publication venues, structured abstracts are introduced as a means to achieve similar clarity and standardization \cite{Budgen2008}, but not in ICSE. Introducing technological rules would, we believe, help in communicating the core of the contribution, both to peer academics and potentially also to industry. Development towards more explicit theory building in software engineering \cite{sjoberg2008building,Stol2015theory} may also pave the way for technological rules as a means to express theoretical contributions.

\subsection{Assessing design knowledge contributions: Rigor, Relevance and Novelty}
Our analysis of rigor, relevance and novelty are based on questions 7-9 in Table~\ref{tab:method-vaquestions}.
\textbf{Rigor} can be considered in terms of the suitability of specific research methods used or in how a certain method is applied. 
Empirical research methods -- quantitative as well as qualitative -- fit well into both problem understanding and validation and we saw examples of very different methods being used.
How rigor is ensured of course depends on the choice of method as we discuss above. 
We found that most authors discussed \textbf{rigor}---not surprising given that these papers were considered as the best papers from an already competitive publishing venue (see RQ3). Whether rigor was discussed for the steps of problem understanding, solution design and solution validation, depended on the paper cluster.
The \emph{solution validation} and \emph{solution design} papers tended to rigorously benchmark their solution against a code base or other artifacts to demonstrate the merits of the proposed approach.  
We found that validating the solutions in industrial contexts was not common in these two clusters of papers. 
Consequently we also found that \textbf{relevance} in terms of specific stakeholders was not discussed much in these papers (as compared to the \emph{descriptive} or \emph{problem-solution} clusters of papers). 

How \textbf{novelty} was discussed by authors varied greatly depending on the paper cluster but also by the author.  
As the papers did not explicate the technological rules, none of them discussed their contribution in terms of technological rule novelty. 
Descriptive papers tended to focus on novelty of the described problem or phenomenon, while problem-solution and solution-design papers focused on novelty of the solution, and the solution-validation emphasized the solution (if refined or new) and the validation insights.  

\section{Recommendations for software engineering research}
\label{sec:recommendations}
 As researchers (and reviewers) ourselves we find that contributions from research papers are often not evident, and thus interested researchers and reviewers may miss the value in the papers. Furthermore, the technological rules, even for papers that aim at producing these rules, are not always easy to identify.  
 To help in the design of research and perhaps also in the review of papers, we suggest using the design lens as follows: 

\begin{itemize}
\item \emph{Explicate design science constructs}: We found design science constructs in most papers, but presenting each of the constructs explicitly, e.g., through the visual abstract~\cite{StoreyESEM17}, could help in communicating 
the research contributions to peer researchers and reviewers. Expressing the technological rules clearly and at a carefully selected level of abstraction, help in communicating the \textbf{novelty} of the contributions and may help in advancing the research in ``standing on each others shoulders''.  
\item \emph{Use real problem instances}: Anchoring research in real problem instances could help to ensure the \textbf{relevance} of the solution. Without reference to an explicit problem instance, the research is at risk of losing the connection with the original question, as the details of a particular intervention are described or assessed by others.
\item \emph{Choose validation methods and context}:
\textbf{Rigor} in terms of method choice is an important consideration.
The choice of methods and context for the validation may be different, depending on the intended scope of the theoretical contribution (i.e., the technological rule).
If the scope is focused on fine tuning the design of an intervention, stakeholders may not need to be directly involved in the validation. 
If however the scope includes the perspective of 
 stakeholders and their context, then methods and study contexts should reflect these perspectives. 
\item \emph{Use the design science lens as a research guide}: The visual abstract and its design science perspective may also be used to guide the design of studies and research programs, i.e., setting particular studies in a particular context. Similarly, the design science perspective can be used as an analysis tool in mapping studies~\cite{Petersen2008}, to assess existing research and identify research gaps that can be explored in future research studies and lead to \textbf{novel} contributions.
\item \emph{Consider research design as a design science}: 
The cluster of \emph{meta} studies, which are primarily aimed for researchers as the stakeholders, indicate that the design science lens also fits for the design and conduct of studies that focus on understanding our research methodology and methods. Papers that address problems in conducting research and propose solutions to help achieve higher quality research contributions are important contributions for our community to reflect and grow in research maturity. Conducting and presenting these in the same way as studies in the software engineering domain adds to their credibility and emphasizes how they are \textbf{relevant} to our community. 
This paper is also an example of a meta study aimed at our research community members as stakeholders.  We created a visual abstract for this paper as well, and it may be found with our online materials (at dsse.org).
\end{itemize}

We hypothesize that following these recommendations, based on our in depth analysis of ICSE distinguished papers, would enable a more consistent assessment of rigor, relevance and novelty of the research contributions, and thus also help the peer review process for future conferences and journals.

\section{Limitations}
\label{sec:limitations}

In order to understand how design science can be a useful lens for describing software engineering research, we considered all papers that have received a distinguished paper award over a five year period within a major venue such as ICSE. We felt these may represent papers that our community considers relevant fine exemplars of SE research. We acknowledge that we would likely see a different result 
for a  different population of papers (e.g., all papers presented at ICSE or in other venues or journals).  That said, we purposefully selected this sample of papers as a exploratory step in our research and don't claim our result would generalize.

Our view of design science may not match other views that are reported in the literature. We developed our view from examining several interpretations of design science as discussed in~\cite{StoreyESEM17} and in Section~\ref{sec:background}.   
Our view was developed over the course of two years spent reading and discussing many design science papers; our interpretation was developed in an iterative manner.  
We have used our visual abstract template in several workshops (notably at ISERN 2017~\footnote{http://www.scs.ryerson.ca/eseiw2017/ISERN/index.html}, RET 2017~\footnote{\url{https://dl.acm.org/citation.cfm?id=3149485.3149522}} and CBSoft 2019~\footnote{\url{https://github.com/margaretstorey/cbsoft2019tutorial}}) and received favorable feedback about the viable application of the template to software engineering papers that contain design knowledge. 

We recognize that our interpretations of the research contributions from the papers we examined may not be entirely accurate or complete.  For this reason we requested feedback from the authors of a selected set of papers to check that our view of the design knowledge in their papers was accurate based on our understanding of their work. Among the responses (7 of 14 of the selected set of paper authors responded), all but one agreed with our summaries presented through the visual abstracts, while this sole initial disagreement was due to misinterpretation of the visual abstract template. This feedback served to some extent as preliminary validation that we were proceeding in the right direction. Consequently, we decided to rely on our judgment for the remaining  papers. 

To do so, we divided papers equally among all the authors assigning two to each paper. They would independently answer the design science questions (as mentioned in Section~\ref{sec:method}), then refer back to the paper in cases of disagreement, and merge our responses until we reached full agreement. Following cases of existing disagreement, we sought additional expert opinions. Finally, we reviewed all of the abstracts as a group to reconfirm our interpretation.  These abstracts are available online and open for external audit by the authors or others in the community. 

To derive clusters of the papers, we followed quite a rigorous process. We met face to face in a several hour workshop and followed up in several sessions over several months to derive the clusters and categorize and reconfirm the categorization of the papers.  
We recognize that how we clustered the papers is potentially subjective and others may feel papers belong in different clusters, and may also find different clusters. 
We have posted all of the visual abstracts and our cluster diagram online which links to all of the visual abstracts (see \url{dsse.org}. We welcome comments on our clusters and the categorization of individual papers.   


\section{Related work}
\label{sec:related_work}

In this paper, we introduced our conceptualization of design science and the visual abstract template, which instantiates our conceptualization and was designed to support communication and dissemination of design knowledge. Furthermore, we reviewed a cohort of software engineering research papers through this lens to investigate its usefulness in the software engineering context. 
In this section of the paper, we extend the scope of related work to include other conceptualizations of design science, as well as other reviews of design science research conducted in a related field. 

Design science has been conceptualized by Wieringa in software engineering~\cite{wieringa_design_2009} and by several researchers in other disciplines, such as information systems~\cite{hevner_design_2004,gregor_positioning_2013,johannesson_introduction_2014} and organization and management~\cite{van_aken_management_2005}. Wieringa describes design science as an act of producing knowledge by designing useful things~\cite{wieringa_design_2009} and makes a distinction between knowledge problems and practical problems. Similarly, Gregor and Hevner emphasize the dual focus on the artifact and its design~\cite{gregor_positioning_2013} in information systems, and argue for an iterative design process where evaluation of the artifact provides feedback to improve both the design process and the artifact. 

In this paper, we do not distinguish between knowledge problems and solution problems within the design sciences but stress that the researcher's task is always to produce knowledge, which in turn can be used by practitioners for solving their problems. Such knowledge may be embedded in artefacts such as tools, models and techniques or distilled to simple technological rules. In line with van Aken~\cite{van_aken_management_2005},  we distinguish between the explanatory sciences and the design sciences as two different paradigms producing different types of theory (explanatory and prescriptive respectively) with different validity criteria. This is similar to  Wieringa's distinction between knowledge problems and practical problem~\cite{wieringa_design_2009}. In our study, we identified one cluster of software engineering papers belonging to the explanatory sciences ('descriptive') and three clusters of papers belonging to the design sciences ('problem-solution', 'solution-design' and 'solution evaluation')

In the management domain, van Aken propose to distinguish management theory, that is prescriptive, from organizational theory, that is explanatory~\cite{van_aken_management_2005}. A corresponding division of software engineering theory has not been proposed yet, although theory types are discussed by software engineering researchers~\cite{sjoberg2008building,stol_uncovering_2013}.

In the literature, design science has been studied and is thereby conceptualized in a number literature studies, which are relevant for this study. 
In the area of information systems, several literature reviews were conducted of design science research. 
Indulska and Recker \cite{Indulska} analyzed design science articles from 2005--07 from well-known information systems conferences. 
They identified 142 articles, which they divided into groups, such as methodology- and discussion-oriented papers and papers presenting implementations of the design science approach. 
They found an increasing number of design science papers over the studied years. 

Deng~\emph{et al.} \cite{Deng17,Deng18} have also published a systematic review of design science articles in information systems. They identified articles by searching in top information systems journals and conferences from the years 2001--15, filtering the results and applying snow-balling, resulting in a final review sample of 119 papers or books. 
In their review, they analyze the topic addressed, artifact type, and evaluation method used. In our review we have classified papers along another dimension, i.e., what types of software engineering design science contributions the papers  present in terms of problem understanding, solution design and solution validation. To our knowledge no reviews of software engineering literature have been made from a design science perspective before.

Wieringa~\emph{et al.}~\cite{Wieringa_ESEM11} have analyzed reasons for the low use of theories in software engineering by studying a set of papers identified in Hannay~\emph{et al.} \cite{Hannay2007}. They compare identified theories in software engineering to general theories with respect to 
level of generalization,
form of theory, and
use of theory,
and argue that the reasons for low use of theories have to do with idealizing assumptions, context of software engineering theories, and that statistical model building needs no theories. 

Concerning relevance, Beecham~\emph{et al.}\ communicated with a test group of practitioners \cite{beecham_making_2014} and found that evidence based on experience was seen as most important, and if it was not available in their own organization, they would seek information from similar organizations in the world for insights on global software engineering. They compare typical sources for software engineering researchers and sources where practitioners seek information, and found that the overlap is very small. Similar findings were obtained by Rainer~\emph{et al.} \cite{Rainer03} in a study based on focus groups with practitioners as well as publications. These observations point to the need for presenting research in a way that is useful for practitioners.  This is also discussed by Grigoleit~\emph{et al.} \cite{grigoleit_quest_2015}, who conclude that practitioners assess the usefulness of many artifacts as being too low. This is inline with our findings, where we put forward the design science lens as a means to better communicate prescriptive research contributions in software engineering. That said, we have not evaluated it with practitioners thus far. 

Another attempt to make evidence available to practitioners is presented by Cartaxo~\emph{et al.} \cite{Cartaxo2016}. 
They present the concept of ``evidence briefings'', which is a way to summarize systematic literature reviews in a one-page format. 
They used accepted information design principles to design the structure of the one-page briefing. The format and content were positively validated by both practitioners and researchers. While evidence briefings may provide an effective way to synthesize evidence from several studies our visual abstract template provides a means to effectively summarize the contribution of one study or research program from a design science perspective.


\section{Conclusions and future work}
\label{sec:conclusion}

Design Science, although suggested for some time as a useful research paradigm for software engineering research, is not commonly used as a way to frame software engineering research contributions.  
Yet our analysis of 38 ICSE distinguished papers indicates that many of these papers can be expressed in terms of the design science paradigm.
Much software engineering research is solution oriented, providing design knowledge, although it is less clear which problems some papers aim to solve.  
 
The technological rule, as a condensed summary of the design knowledge, offers a means to communicate not just the solutions designed or validated, but also the problems addressed. 
We were able to derive technological rules from most papers, although they were not explicitly stated as such in these papers. 
In future work, we aim to investigate how technological rules could be linked across different research contributions that address the same underlying problem.  
A higher level technological rule could be decomposed into more narrow but related rules, thus bringing insights across multiple papers that are linked by context, intervention type and effect.  
Currently, we lack the machinery in our community to link papers at this theoretical level and the results in papers remain as silos and are often not even referenced in related work. 
The technological rule template could help fill this gap and help us to better understand what we know and what we don't know yet about certain problems and challenges in software engineering.

Also as future work, we wish to investigate if the design science visual abstract (or some variant of it) could provide an efficient way to present software engineering research contributions to industry.  We published our abstracts from this study online---but it remains to be seen if industry finds this format useful or not.
We expect that extracting technological rules from a set of papers that address a common problem or topic is likely to be of more value to industry (this was not a goal of this current work). 
In the meantime, we anticipate that our analysis of ICSE distinguished papers through the design science lens, may help our community increase adoption of the design science lens, which we anticipate in turn will allow us to do a better job of communicating, understanding and building on each others' work. 

Furthermore, as a means for spreading the word of this research to the community,  it is our intention to contact editors of important journals as well as  program chairs of relevant conferences such as ICSE and promote the adoption of VAs for authors that submit a research paper.  

\begin{acknowledgements}
We would like to thank Cassandra Petrachenko for her careful edits of our paper. 
Daniela Soares Cruzes, Johan Lin{\aa}ker, Sergio Rico and Eirini Kalliamvakou gave us helpful comments on an earlier draft of this paper. 
We would also like to thank some of the authors of ICSE distinguished papers for giving us feedback, as well as participants at ISERN 2017, RET 2017, and CBSoft 2019 for trying out the visual abstract. 
Their enthusiasm encouraged us to continue with this research. 
We thank the anonymous EMSE reviewers for helping us sharpening the contribution of this paper.
The research was partially funded by the Faculty of Engineering at Lund University through the Lise Meitner guest professorship (Storey), the ELLIIT strategic research area (Engstr\"om), and the EASE industrial excellence center (Runeson).
\end{acknowledgements}

\bibliographystyle{spmpsci}      
\bibliography{references}
\nociteI{*}
\bibliographystyleI{spmpsci}
\bibliographyI{icserefs}



%
%

\end{document}